\documentclass[Afour,sageh,times]{sagej}
\usepackage{moreverb,url}
\usepackage[colorlinks,bookmarksopen,bookmarksnumbered,citecolor=red,urlcolor=red]{hyperref}

\usepackage{multirow}
\usepackage{amssymb}
\usepackage{amsmath}
\usepackage{graphicx}
\usepackage{algorithmic}
\usepackage{algorithm}
\usepackage{color,graphicx,bm,multirow}
\renewcommand{\t}[1]{\textrm{\scriptsize #1}}

\newcommand\BibTeX{{\rmfamily B\kern-.05em \textsc{i\kern-.025em b}\kern-.08em

T\kern-.1667em\lower.7ex\hbox{E}\kern-.125emX}}

\def\volumeyear{2016}

\begin{document}
\runninghead{Xinming Qin, Honghui Shang, Lei Xu, Wei Hu, Jinlong Yang, Shigang Li, ,Yunquan Zhang}

\title{The static parallel distribution algorithms for hybrid density-functional calculations in HONPAS package}
\author{Xinming Qin$^{2\dagger}$,Honghui Shang*$^{1\dagger}$,Lei Xu$^1$, Wei Hu$^2$, Jinlong Yang$^2$, Shigang Li$^1$, Yunquan Zhang$^1$}

\affiliation{\affilnum{1}State Key Laboratory of Computer Architecture, Institute of Computing Technology, Chinese Academy of Sciences, Beijing, China \\
\affilnum{2}Hefei National Laboratory for Physical
Sciences at Microscale, Department of Chemical Physics, and
Synergetic Innovation Center of Quantum Information and Quantum
Physics, University of Science and Technology of China, Hefei, Anhui
230026, China }

\corrauth{*Corresponding author: E-mail: shanghui.ustc@gmail.com\\ $\dagger$ Both authors contributed equally to this work.}

\begin{abstract}

Hybrid density-functional calculation is one of the most commonly adopted electronic structure theory used in computational chemistry and materials science because of its balance between accuracy and computational cost. Recently, we have developed a novel scheme called NAO2GTO to achieve linear scaling (Order-N) calculations for hybrid density-functionals\cite{Shang-JCP}. In our scheme, the most time-consuming step is the calculation of the electron repulsion integrals (ERIs) part. So how to create an even distribution of these ERIs in parallel implementation is an issue of particular importance. Here, we present two static scalable distributed algorithms for the ERIs computation. Firstly, the ERIs are distributed over ERIs shell pairs. Secondly, the ERIs is distributed over ERIs shell quartets. In both algorithms, the calculation of ERIs is independent of each other, so the communication time is minimized. We show our speedup results to demonstrate the performance of these static parallel distributed algorithms in the Hefei Order-N packages for \textit{ab initio} simulations (HONPAS)\cite{Qin2015}. 

\end{abstract}

\maketitle

\section{INTRODUCTION}
The electronic structure calculations based on density functional theory (DFT) \cite{Parr, DFT_HK, DFT_KS} are the work-horse of computational chemistry and materials science. However, widely used semi-local density functionals could underestimate the band gaps because of its inclusion of the unphysical self-interaction \cite{Cohen-Yang}. A possible solution is to add the nonlocal Hartree-Fock type exchange (HFX) into semi-local density-functionals to construct hybrid functionals\cite{Becke_1993,Stephens_1994,Janesko_2009,Paier_2009,HSE-better-1,HSE-better-2,HSE-better-3,HSE03-1,HSE03-2,HSE06,Gaussian,VASP-HSE}. However, the drawback of hybrid density-functionals is that it is significantly more expensive than conventional DFT. The most time-consuming part in hybrid density-functional calculations becomes construction of HFX matrix, even with the appearance of fast linear scaling algorithms that overcome the bottlenecks encountered in conventional methods\cite{Schwegler_1996, Burant_1996,Schwegler_1997, Ochsenfeld_1998,Polly_2004,Tymczak_2005,Sodt_2008,Guidon_2010,Merlot_2014}. As a result, hybrid density-functional calculations must make efficient use of parallel computing resources in order to reduce the execution time of HFX matrix construction. 

The implementation of hybrid density-functionals for solid state physics calculations are mostly based on plane waves~(PW)\cite{Gonze2002,Gonze2016,VASP-HSE} or linear combination of atomic orbitals~(LCAO)\cite{HSE06,Gaussian,CRYSTAL} method. The atomic orbitals basis set is efficient for real-space formalisms, which have attracted considerable interest for DFT calculations because of their favorable scaling with respect to the  number of atoms and their potential for massively parallel implementations for large-scale calculations~\cite{DMOL,SIESTA,Blum2009,Havu2009,Ren2012,GAPW,Mohr2014,Gaussian,Shang-JCP}. Unlike plane wave method, when constructing HFX matrix within LCAO method, we must first calculate the ERIs via the atomic orbitals. There are currently two types of atomic orbits that are most commonly used. The first is gaussian type orbital(GTO), as adopted in Gaussian\cite{Gaussian} and CRYSTAL\cite{CRYSTAL}, its advantage is to calculate ERIs analytically. The second is numerical atomic orbital (NAO), which is adopted in SIESTA\cite{SIESTA}, DMOL\cite{DMOL}, OPENMX\cite{OPENMX}, et al.. The advantage of NAO is its strict locality, which naturally leads to lower order scaling of computational time versus system size. In order to take advantages of both types of atomic orbitals, we have proposed a new scheme called NAO2GTO\cite{Shang-JCP}, in which GTO can be used for analytical computation of ERIs in a straightforward and efficient way, while NAO can be employed to set the strict cutoff for atomic orbitals. After employing several ERI screening techniques, the construction of HFX matrix can be very efficient and scale linearly\cite{Shang-JCP,Qin2015}.


Parallelization of HFX matrix construction faces two major problems of load imbalance and high communication cost. The load imbalance arises from the irregularity of the independent tasks available in the computation, which is due to the “screening” procedure and different types of shell quartets distributed among processes. The high communication cost is from interprocessor communication of the density and/or HFX matrices, which is associated with the data access pattern. It is well known that NWChem\cite{NWChem} and CP2K/Quickstep\cite{Quickstep} are the most outstanding softwares in the field of high performance parallel quantum chemical computing, and both of them use GTOs to construct HFX matrix. In NWChem, the parallelization of HFX matrix construction is based on a static partitioning of work followed by a work stealing phase\cite{NWChem_HF1,NWChem_HF2}. The tasks are statically partitioned throughout the set of shell (or atom) quartets, and then the work stealing phase acts to polish the load balance. As a result, this parallel implementation gives very good parallel scalability of Hartree{\textendash}Fock calculations\cite{NWChem_HF1}. In CP2K/Quickstep, the HFX parallelization strategy is to replicate the global density and HFX matrix on each MPI process in order to reduce communication. A load balance optimization based on simulated annealing and a binning procedure to coarse grain the load balancing problem have been developed\cite{CP2K-2}. However, this approach may limit both system size and ultimately parallel scalability. 


As the ERIs calculation is the most computationally demanding step in the NAO2GTO scheme, the development of the new parallel algorithms is of particular importance. Previously, for codes using localized atomic orbitals, the parallelization of ERIs are mainly implemented to treat finite, isolated systems~\cite{Schmidt1993,ALEXEEV2002,NWChem_HF1,
NWChem_HF2}, but only a few literature reports exist for the treatment of periodic boundary conditions with such basis sets~\cite{Bush2011,CP2K-2}, in which the Order-N screening for the ERIs calculations has not been considered. The purpose of this work is to present the static parallel distribution algorithms for the NAO2GTO scheme\cite{Shang-JCP} with Order-N performance in HONPAS code\cite{Qin2015}. In our approaches, the calculations of ERIs are not replicated, but are distributed over CPU cores, as a result, both the memory and the CPU requirements of the ERIs calculation are paralleled. The efficiency and scalability of these algorithms are demonstrated by benchmark timings 
in periodic solid system with 64 silicon atoms in the unit cell.

The outline of this paper is as follows: In Section 2, we begin with a description of the theory of hybrid functionals. In Section 3, we describe the detail implementation of our parallel distribution . In Section 4, we present the benchmark results and the efficiency of our scheme.

\section{Fundamental Theoretical Framework}
Before addressing the parallel algorithms, we recall the basic equations used in this work. A spin-unpolarized 
notation is used throughout the text for the sake of simplicity, but a formal generalization to
the collinear spin case is straightforward. In Kohn-Sham DFT, the total-energy functional is given as 
\begin{equation}
E_\t{KS}= T_\t{s}[n]+E_\t{ext}[n]+E_\t{H}[n]+E_\t{xc}[n] + E_\t{nuc-nuc}\;.
\label{eq:KSTOT}
\end{equation}
Here, $n(\mathbf{r})$ is the electron density, $T_\t{s}$ is the kinetic energy of non-interacting electrons, while $E_\t{ext}$ is external energy stemming from the electron-nuclear attraction, $E_\t{H}$ is the Hartree energy, $E_\t{xc}$ is the exchange-correlation energy, and $E_\t{nuc-nuc}$ is the nucleus-nucleus repulsion energy.

The ground state electron density~$n_0(\mathbf{r})$ (and the associated ground state total energy) 
is obtained by variationally minimizing Eq.~(\ref{eq:KSTOT}) under the constraint that
the number of electrons $N_e$ is conserved. This yields the chemical potential $\mu=\delta E_{KS}/\delta n$ of the electrons
and the Kohn-Sham single particle equations
\begin{equation}
\hat{h}_\t{KS}\psi_i = \left[ \hat{t}_\t{s} + v_\t{ext}(r)+v_\t{H}+v_\t{xc}\right] \psi_i = \epsilon_{p} \psi_i 
\label{eq:ks-equation}
\end{equation}
for the Kohn-Sham Hamiltonian~$\hat{h}_\t{KS}$. In Eq.~(\ref{eq:ks-equation}), $\hat{t}_\t{s}$ denotes the kinetic energy operator, $v_\t{ext}$ the external potential, $v_{H}$ the Hartree potential, and $v_{xc}$ the exchange-correlation potential. Solving Eq.~(\ref{eq:ks-equation}) yields the Kohn-Sham single particle states~$\psi_p$ and their eigenenergies~$\epsilon_{p}$.
The single particle states determine the electron density via
\begin{equation}
n(\mathbf{r})=\sum_i f_i |\psi_i|^2,
\end{equation}
in which $f_i$ denotes the Fermi-Dirac distribution function.

To solve Eq.~(\ref{eq:ks-equation}) in numerical implementations,
the Kohn-Sham states are expanded in a finite basis set. For periodic systems, the crystalline orbital
$\psi_{i}(\mathbf{k,r})$  normalized in all space is a linear combination of Bloch functions
$\phi_{\mu}(\mathbf{k,r})$,
defined in terms of atomic orbitals $\chi_{\mu}^{\mathbf{R}}(\mathbf{r})$.\\
\begin{equation}
\psi_{i}(\mathbf{k,r})=
\sum_{\mu}c_{\mu,i}(\mathbf{k})\phi_{\mu}(\mathbf{k,r})
\end{equation}
\begin{equation}
\phi_{\mu}(\mathbf{k,r})=\dfrac{1}{\sqrt{N}}\sum_{\mathbf{R}}\chi_{\mu}^{\mathbf{R}}(\mathbf{r})\mathbf{e}^{i\mathbf{k}\cdotp
(\mathbf{R+r_{\mu}})}
\end{equation}
where the Greek letter $\mu$ is the index of atomic orbitals, $i$ is
the suffix for different bands, $\mathbf{R}$ is the origin of a unit
cell, N is the number of unit cells in the system.
$\chi_{\mu}^{\mathbf{R}}(\mathbf{r})=\chi_{\mu}(\mathbf{r-R-r_\mu})$
is the $\mu$-th atomic orbital, whose center is displaced from the
origin of the unit cell at $\mathbf{R}$ by $\mathbf{r}_\mu$.
$c_{\mu,i}(\mathbf{k})$ is the wave function coefficient, which is
obtained by solving the following equation,
\begin{equation}
H(\mathbf{k})c(\mathbf{k})=E(\mathbf{k})S(\mathbf{k})c(\mathbf{k})
\end{equation}
\begin{equation}
 [H(\mathbf{k})]_{\mu\nu}=\sum_{\mathbf{R}}H_{\mu\nu}^{\mathbf{R}}\mathbf{e}^{i\mathbf{k}\cdotp (\mathbf{R+r_{\nu}-r_{\mu}})}
\end{equation}
\begin{equation}
 H^{\mathbf{R}}_{\mu\nu}=<\chi_{\mu}^{\mathbf{0}}|\hat{H}|\chi_{\nu}^{\mathbf{R}}>
 \label{eq:Huv}
\end{equation}
\begin{equation}
 [S(\mathbf{k})]_{\mu\nu}=\sum_{\mathbf{R}}S_{\mu\nu}^{\mathbf{R}}\mathbf{e}^{i\mathbf{k}\cdotp (\mathbf{R+r_{\nu}-r_{\mu}})}
\end{equation}
\begin{equation}
 S^{\mathbf{R}}_{\mu\nu}=<\chi_{\mu}^{\mathbf{0}}|\chi_{\nu}^{\mathbf{R}}>
\end{equation}
In Eq. (\ref{eq:Huv}), $H^{\mathbf{R}}_{\mu\nu}$ is a matrix element
of the one-electron Hamiltonian operator $\hat{H}$ between the
atomic orbital $\chi_{\mu}$ located in the central unit cell $0$ and
$\chi_{\nu}$ located in the unit cell $\mathbf{R}$.

It should be noted that, the exchange-correlation potential $v^{xc}$ is a local and periodic in semi-local DFT, 
while in Hartree-Fock and hybrid functionals, the Hartree-Fock exchange~(HFX) potential matrix element is defined as:
\begin{equation}
  [V^{X}]_{\mu\lambda}^{\mathbf{G}}=-\frac{1}{2}\sum_{\nu\sigma}\sum_{\mathbf{N,H}}P_{\nu\sigma}^\mathbf{H-N}\mathbf{[(\chi_{\mu}^{0}\chi_{\nu}^{N}|\chi_{\lambda}^{G}\chi_{\sigma}^{H})]}
  \label{eq:vx}
\end{equation}
where $\mathbf{G}$, $\mathbf{N}$, and $\mathbf{H}$ represent
different unit cells. The density matrix element
$P_{\nu\sigma}^\mathbf{N}$ is computed by an integration over the
Brillouin zone (BZ)
\begin{equation}
 P_{\nu\sigma}^{\mathbf{N}}=\sum_j \int_{BZ} c_{\nu,j}^{*}(\mathbf{k})c_{\sigma,j}(\mathbf{k}) \theta(\epsilon_F-\epsilon_j(\mathbf{k}))\mathbf{e}^{i\mathbf{k}\cdotp \mathbf{N}}d\mathbf{k}
\end{equation}
where $\theta$ is the step function, $\epsilon_F$ is the fermi
energy and $\epsilon_j(\mathbf{k})$ is the $j$-th eigenvalue at
point $\mathbf{k}$.

In order to calculate the following ERI in Eq. (\ref{eq:vx})
\begin{equation}
 \mathbf{(\chi_{\mu}^{0}\chi_{\nu}^{N}|\chi_{\lambda}^{G}\chi_{\sigma}^{H})=\int\int   \frac{\chi_{\mu}^{0}(r)\chi_{\nu}^{N}(r)\chi_{\lambda}^{G}(r')\chi_{\sigma}^{H}(r')}{|r-r'|}
 drdr'}
\end{equation}
we use NAO2GTO scheme described in the following section.

\begin{figure}
\includegraphics[width=0.5\textwidth]{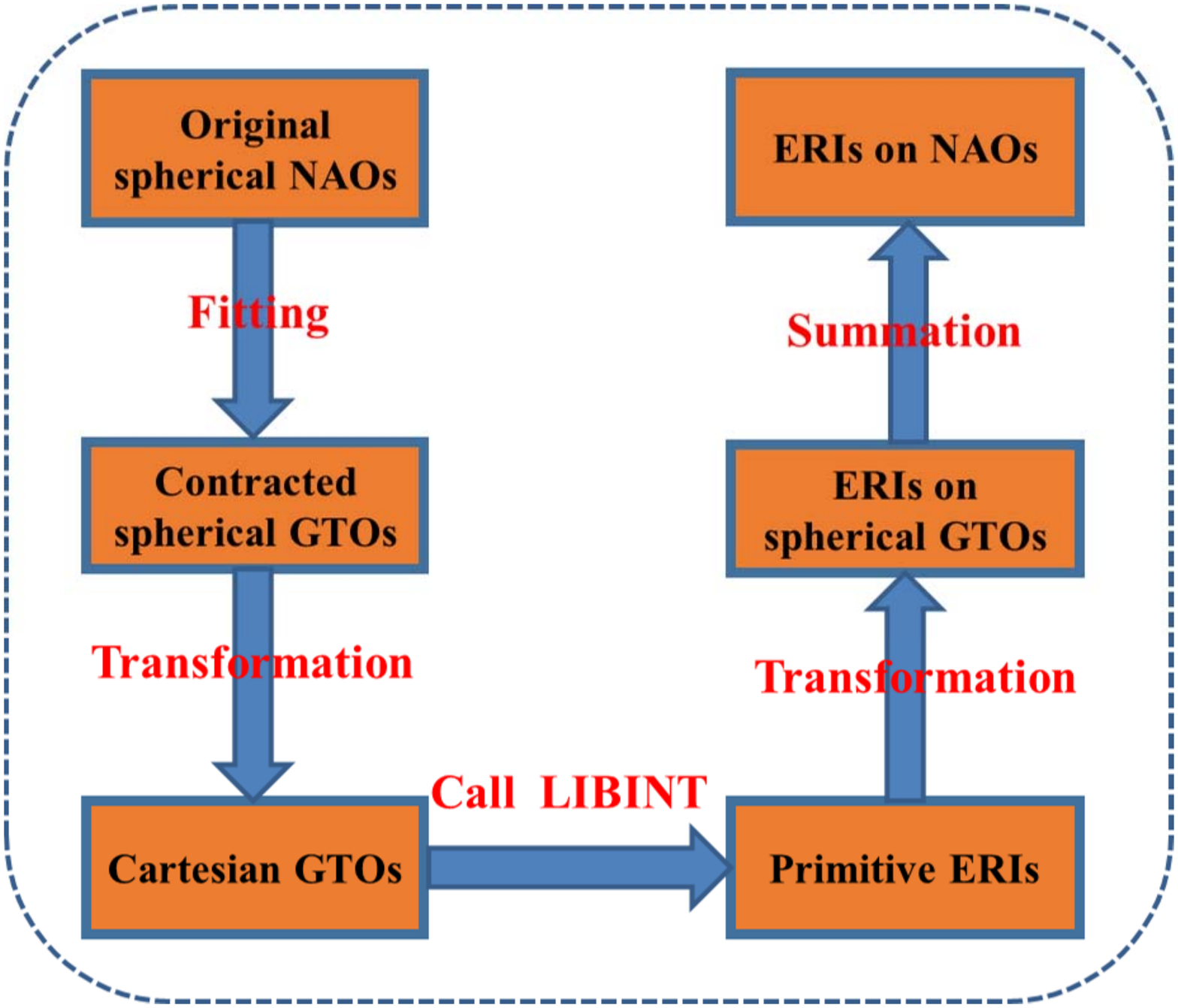}
\caption{The flowchart of the NAO2GTO scheme in HONPAS.}
\label{fig:honpas_ERI}
\end{figure}

Following the flowchart in Fig.\ref{fig:honpas_ERI}, the NAO2GTO scheme is to firstly fit the NAO with GTOs, and then to calculate ERIs analytically with fitted GTOs.
A NAO is a product of a numerical radial function and a spherical harmonic
\begin{equation}
 \phi_{Ilmn}(\mathbf{r})=\varphi_{Iln}(r)\mathbf{Y}_{lm}(\hat{r})
\end{equation}
The radial part of the numerical atomic orbital $\varphi_{Iln}(r)$
is calculated by the following equation:
\begin{equation}
 (-\frac{1}{2}\mathbf{\frac{1}{r}}\frac{\mathrm{d^{2}}}{\mathrm{d}r^{2}}r+\frac{l(l+1)}{2r^{2}}+
 V(r)+V_{cut})
\varphi_{Iln}(r)=\epsilon_{l}\varphi_{Iln}(r)
\end{equation}
where $V(r)$ denotes the electrostatic potential for orbital
$\varphi_{Iln}(r)$, and $V_{cut}$ ensures a smooth decay of each
radial function which is strictly zero outside a confining radius
$r_{cut}$.

\section{METHODS}
\section{NAO2GTO scheme}

In our NAO2GTO approach, the radial part of
 the NAO $\varphi_{Iln}(r)$ is fitted by the summation of several GTOs,
 denoted as $\chi(r)$
\begin{equation}
\chi(r)\equiv\sum_m{D_{m} r^{l} \exp(-\alpha_{m}r^{2})}
\end{equation}
Parameters $\alpha_{m}$ and $D_{m}$ are determined by minimizing the
residual sum of squares of the difference
\begin{equation}
 \sum_i{(\chi(r_i)/r_i^{l}-\varphi_{Iln}(r_i)/r_i^{l})^2}
\end{equation}
 
In practice of the solid system calculation, too diffused basis set may cause
convergence problem\cite{CP2K-2}, as a result the exponents smaller than 0.10 are usually not needed,
and we made a constraint, i.e. ($\alpha > 0.1$) during our minimal search.
First we use constrained genetic algorithm\cite{global-min-1,global-min-2} to do a global search for initial guess and then do a constrained local minimal search using trust-region-reflective algorithm, which is a subspace trust-region method and is based on the interior-reflective Newton method described in \cite{local-min-1} and \cite{local-min-2}. Each iteration involves the approximate solution of a large linear system using the method of preconditioned conjugate gradients. We make N ($N>500$) global searches to make sure to have a global minimal. 

\begin{algorithm}
\caption{The algorithm of NAO2GTO fitting scheme.}
\begin{algorithmic}
\FOR{$iter=1$ to $N$}
   \STATE constrained genetic algorithm get initial $\alpha^{iter}_{m}$ and $D^{iter}_{m}$ \STATE constrained local minimal search to get $\alpha^{iter}_{m}$ and $D^{iter}_{m}$ 
   \STATE $err=\sum_i{[ \sum_m{D_{m} \exp(-\alpha_{m}r_i^{2})}  -\varphi_{Iln}(r_i)/r_i^{l}]^2}$
   \IF{$iter=1$ .or. $err<best\_err$ }
   \STATE $best\_err=err$ 
   \STATE $\alpha_{m}=\alpha^{iter}_{m}$ and $D_{m}=D^{iter}_{m}$ 
   \ENDIF
\ENDFOR
\end{algorithmic}
\label{algo:minimal-search}
\end{algorithm}

Using the above fitting parameters, we build all ERIs that needed for the HFX. 
In our implementation, the full permutation
symmetry of the ERIs has been considered for solid systems:
\[
 (\mu^{\mathbf{0}}\nu^{\mathbf{H}}|\lambda^{\mathbf{G}}\sigma^{\mathbf{N}})=
   (\mu^{\mathbf{0}}\nu^{\mathbf{H}}|\sigma^{\mathbf{N}}\lambda^{\mathbf{G}})=
\]
\[
   (\nu^{\mathbf{0}}\mu^{\mathbf{-H}}|\lambda^{\mathbf{G-H}}\sigma^{\mathbf{N-H}})=
   (\nu^{\mathbf{0}}\mu^{\mathbf{-H}}|\sigma^{\mathbf{N-H}}\lambda^{\mathbf{G-H}})=
\]
\[
   (\lambda^{\mathbf{0}}\sigma^{\mathbf{N-G}}|\mu^{\mathbf{-G}}\nu^{\mathbf{H-G}})=
   (\lambda^{\mathbf{0}}\sigma^{\mathbf{N-G}}|\nu^{\mathbf{H-G}}\mu^{\mathbf{-G}})=
\]
\begin{equation}
   (\sigma^{\mathbf{0}}\lambda^{\mathbf{G-N}}|\mu^{\mathbf{-N}}\nu^{\mathbf{H-N}})=
   (\sigma^{\mathbf{0}}\lambda^{\mathbf{G-N}}|\nu^{\mathbf{H-N}}\mu^{\mathbf{-N}})
\end{equation}
In this way, we save a factor of 8 in the number of integrals that have to be calculated.

When calculating the ERIs with GTOs, the atomic orbitals
are grouped into shells according to the angular momentum.
The list need to be distributed in parallel is in fact the shell quartet. For a shell with a angular momentum of $l$, the number of atomic orbital basis functions is $2l+1$, so in a shell quartet integral $(IJ|KL)$, we calculate in total $(2l_I+1)(2l_J+1)(2l_K+1)(2l_L+1)$ atomic basis orbital ERIs together.  As a result, the computational expense 
is strongly dependent on the angular momenta of the shell quartet. It is a challenge to distribute these shell ERIs
not only in number but also considering the time-weight.

In our NAO2GTO scheme, two shell pair lists (list-IJ and list-KL) are firstly preselected according to Schwarz screening\cite{Schwarz}, 
\begin{equation}
|(\mu\nu|\lambda\sigma)| \leqslant \sqrt{(\mu\nu|\mu\nu)(\lambda\sigma|\lambda\sigma)}
\label{eq:schwarz}
\end{equation}
and only the shell list indexes with $(IJ|IJ)>\tau$ or $(KL|KL)>\tau$ (here $\tau$ is the drop tolerance) are stored.
As shown in Eq. (\ref{eq:vx}), the first index I runs only within the unit cell, while the indexes (J,K,L) run over the whole supercell, so the list-IJ is smaller than the list-KL. Then in the ERIs calculations, the loops run over these two shell lists.

Then before the calculation of every ERI, we use Schwarz inequality Eq. (\ref{eq:schwarz}) again to
estimate a rigorous upper bound, that only the ERIs with non-negligible contributions are calculated,
we note this screening method as Schwarz screening. Using the exponential decay of the charge distributions, the Schwarz screening reduces the total number of ERIs to be computed from $O(N^4)$ to $O(N^2)$. The Schwarz screening tolerance is set to 10$^{-5}$ in the following calculation. 
In addition, the NAO screening is also adopted as the NAO is strictly truncated\cite{Shang-JCP}. The NAO screening is safe in the calculation of the short-range ERI because 
in this case the Hartree-Fock exchange~(HFX) Hamiltonian matrix is sparse due to the screened Coulomb potential\cite{Izmaylov2006}. As a result, we store this HFX Hamiltonian with a sparse matrix data structure.

In practice, it also should be noted that as the angular part of the NAOs is spherical harmonic while the GTOs are Cartesian Gaussians, we need to make a transformation between Cartesian and Spherical harmonic functions. The difference between these two harmonic functions is the number of atomic orbitals including in the shells whose angular momentum are larger than 1.  For example, a d-shell has 5 Spherical orbitals, but have 6 Cartesian orbitals. A detailed transformation method can be find in Ref.\cite{Schlegel}.

\subsection{Parallel schemes}
The ERIs have four indexed that can be paralleled. One possible parallel scheme to make distribution of just one shell index, however, as the index number in one shell is  too small to make distribution over CPU cores, such shell-one distribution may cause serious load imbalance.  

The other parallel scheme is to make distribution for shell-pair (list-KL). It is a straightforward way to parallelize the ERIs as in practice we loop over two pair lists. 
However, although the shell-pair can be distributed evenly before ERIs calculations, ERI-screening is needed during the ERIs calculations, which also causes load imbalance.  
The practical implementation of the described formalism closely follows the flowchart shown in
Algorithm \ref{algo:para-pair}. After the 
building list-KL is completed, we distributed 
it into CPU cores with list-KL-local at every 
cores. Then in the following ERIs screening and
calculation, only loops over list-KL-local is needed for every core. The advantage of this scheme is that it naturally bypasses the the communication process, and every CPU core only
go over and compute its assigned list-KL-local.
However, although the list-KL is distributed evenly over processors, the ERI screening is located after the parallelization, which causes different number of shell ERIs that need to be calculated in every  processor. Such different
number of shell ERIs makes load imbalance.

In order to achieve load balance, distribution of individual shell-quartet$(IJ|KL)$ after the ERI screening process a possible choice. Even if the computational time is nonuniformity in the case of the different shell type, this distribution can also yield an even time over CPU cores because of its smallest distribution chunks. The practical implementation of the shell-quartet algorithm is shown in
Algorithm \ref{algo:para-quartet}. Every 
CPU core will go over the global pair lists(
list-IJ and list-KL), and make the ERI screening
to determine which ERIs are needed to be computed.
Then a global counter is set to count the
number of computed ERIs, this counter is distributed over CPU cores to make sure the number of calculated ERIs is evenly distributed. 
The disadvantage of this algorithm is that every
processor need to make the whole ERI screening, while in the above parallel-pair algorithm, only the ERI screening in its local lists is needed.
Such globally calculated ERIs screening decreases the parallel efficiency.

\begin{algorithm}
\caption{Flowchart of the parallel-pair algorithms for ERIs. Here the shell-pair-list-KL-local means to distribute the shell-list-KL over CPU cores at the beginning. The description of Schwarz screening and NAO screening are in the text.}
\begin{algorithmic}

\STATE get shell-pair-list-KL-local 
 \FOR{list-IJ in shell-pair-list-IJ}
 \FOR{list-KL in shell-pair-list-KL-local} 
       \IF{Schwarz screening.and. NAO screening}
      \STATE  compute shell ERI $(IJ|KL)$
       \ENDIF 
 \ENDFOR
 \ENDFOR
\end{algorithmic}
\label{algo:para-pair}
\end{algorithm}

\begin{algorithm}
\caption{Flowchart of the parallel-quartet algorithms for ERIs. Here N refers to the total number of the CPU cores, current-core refers to the index of the current processor. The description of Schwarz screening and NAO screening are in the text.}
\begin{algorithmic}
 \FOR{list-IJ in shell-pair-list-IJ}
 \FOR{list-KL in shell-pair-list-KL} 
       \IF{Schwarz screening.and. NAO screening}
      \STATE  i ++ 
       \IF{ i mod N eq current-core} 
      \STATE  compute shell ERI $(IJ|KL)$
       \ENDIF 
       \ENDIF 
 \ENDFOR
 \ENDFOR
\end{algorithmic}
\label{algo:para-quartet}
\end{algorithm}

\section{RESULTS AND DISCUSSION}
\begin{figure}
\begin{center}
\begin{tabular}{c c c c }
\hline 
     & basis set & shells & NAOs  \\
     \hline
Si64 & SZ        &   128     & 256       \\
Si64 & DZP       &   320     & 832      \\
\hline 
\end{tabular}
\end{center}
\includegraphics[width=0.5\textwidth]{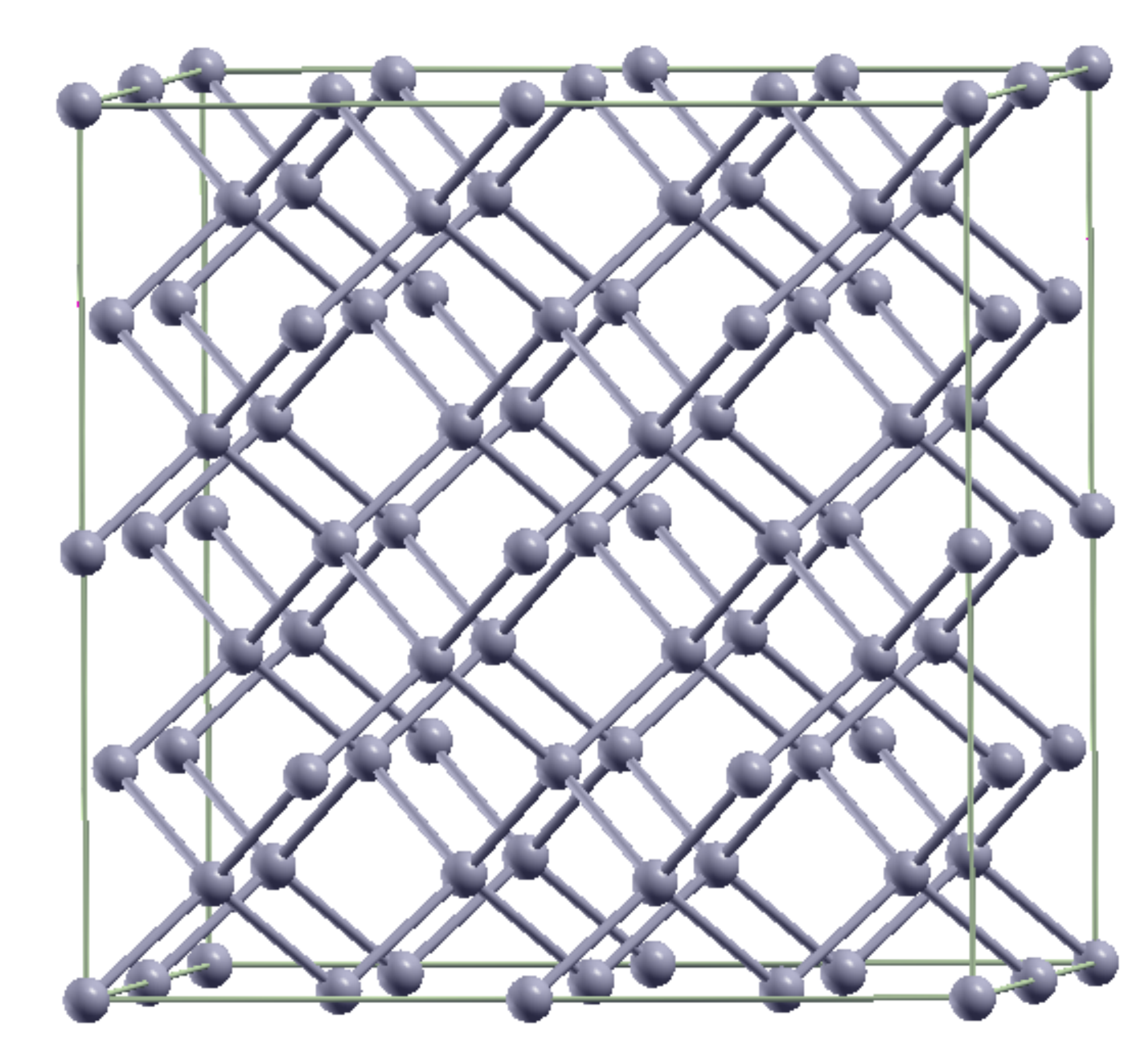}
\caption{The silicon bulk contained 64 atoms in the unit cell that used as benchmark system in this work. In the upper table, the number of shells as well as the number of the NAO basis functions
for different basis sets are listed.}
\label{fig:si_64atoms}
\end{figure}

In order to demonstrate the performance of the 
above two static parallel schemes, we use silicon bulk contained 64 atoms in the unit cell
as a test case as shown in Fig. \ref{fig:si_64atoms}. Norm-conserving pseudopotentials generated with the Troullier-Martins\cite{TM} scheme, in fully separable form
developed by Kleiman and Bylader\cite{KB}, are used to represent
interaction between core ion and valence electrons.  The screened hybrid
functional HSE06\cite{HSE06} was used in the following calculations.  Both single-zeta~(SZ) contained s and p shells and double-zeta plus polarization~(DZP) basis set contained s,p and d shells are considered.  All calculations were carried out on Tianhe-2 supercomputer located in National Supercomputer Center in Guangzhou, China, the configuration of the machine is shown in Table \ref{tab:tianhe}. The Intel Math Kernel Library(version 10.0.3.020) is used in the calculations.

\begin{table}
\caption{The parameters for each node of Tianhe-2.}
\label{tab:tianhe}
\begin{tabular}{c c}
\hline
Component &  Value\\
\hline
CPU  &  Intel(R) Xeon(R) CPU E5-2692 \\
Freq.~(GHz) &  2.2 \\
Cores  & 12 \\
\hline
\end{tabular}
\end{table}

\begin{figure}
\includegraphics[width=0.5\textwidth]{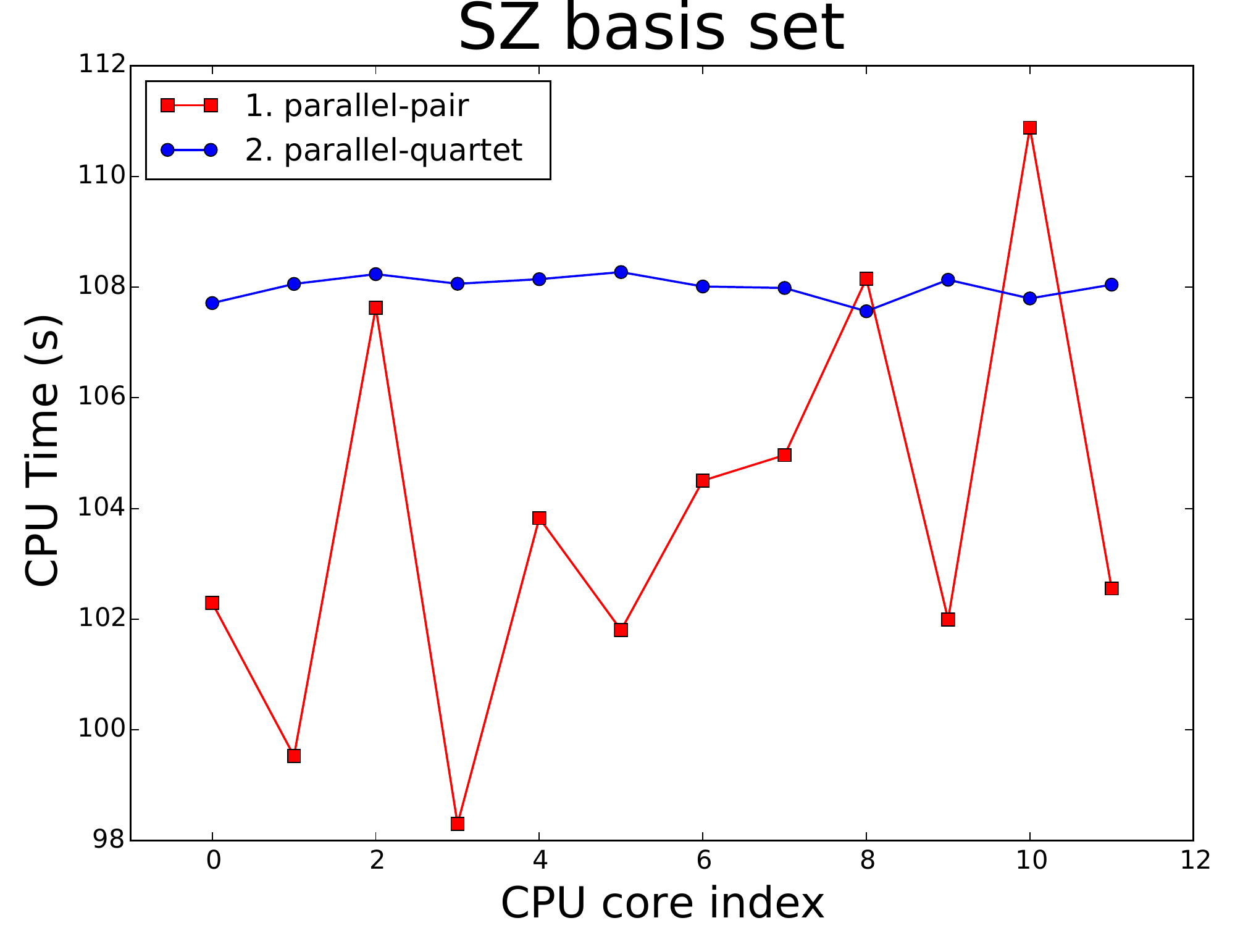}
\caption{The load balance for bulk silicon supercell with 64 atoms using SZ basis set at 12 CPU cores.}
\label{fig:si_64atoms_sz_12core}
\end{figure}

\begin{figure}
\includegraphics[width=0.5\textwidth]{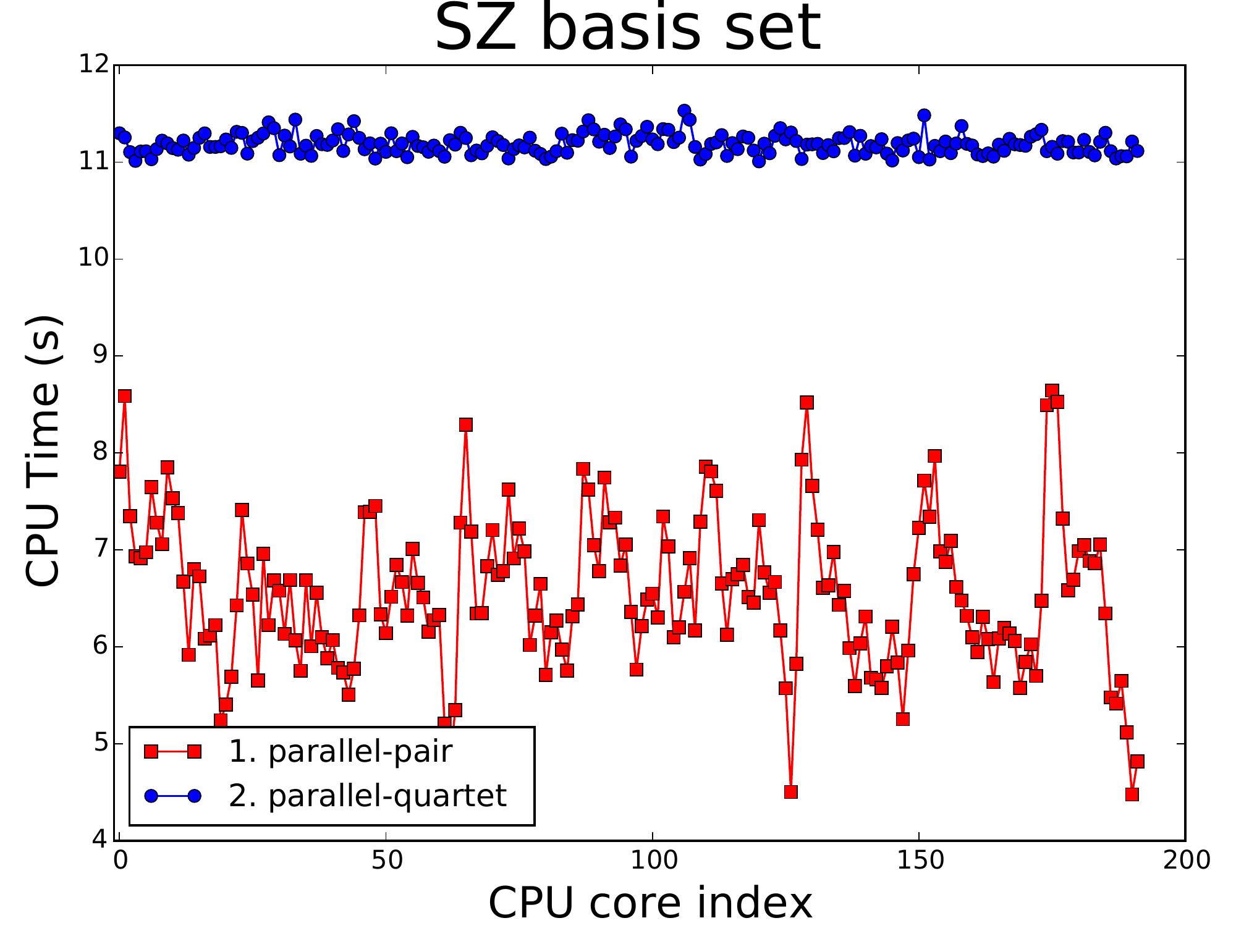}
\caption{The load balance for bulk silicon supercell with 64 atoms using SZ basis set at 192 CPU cores.}
\label{fig:si_64atoms_sz_192core}
\end{figure}

\begin{figure}
\includegraphics[width=0.5\textwidth]{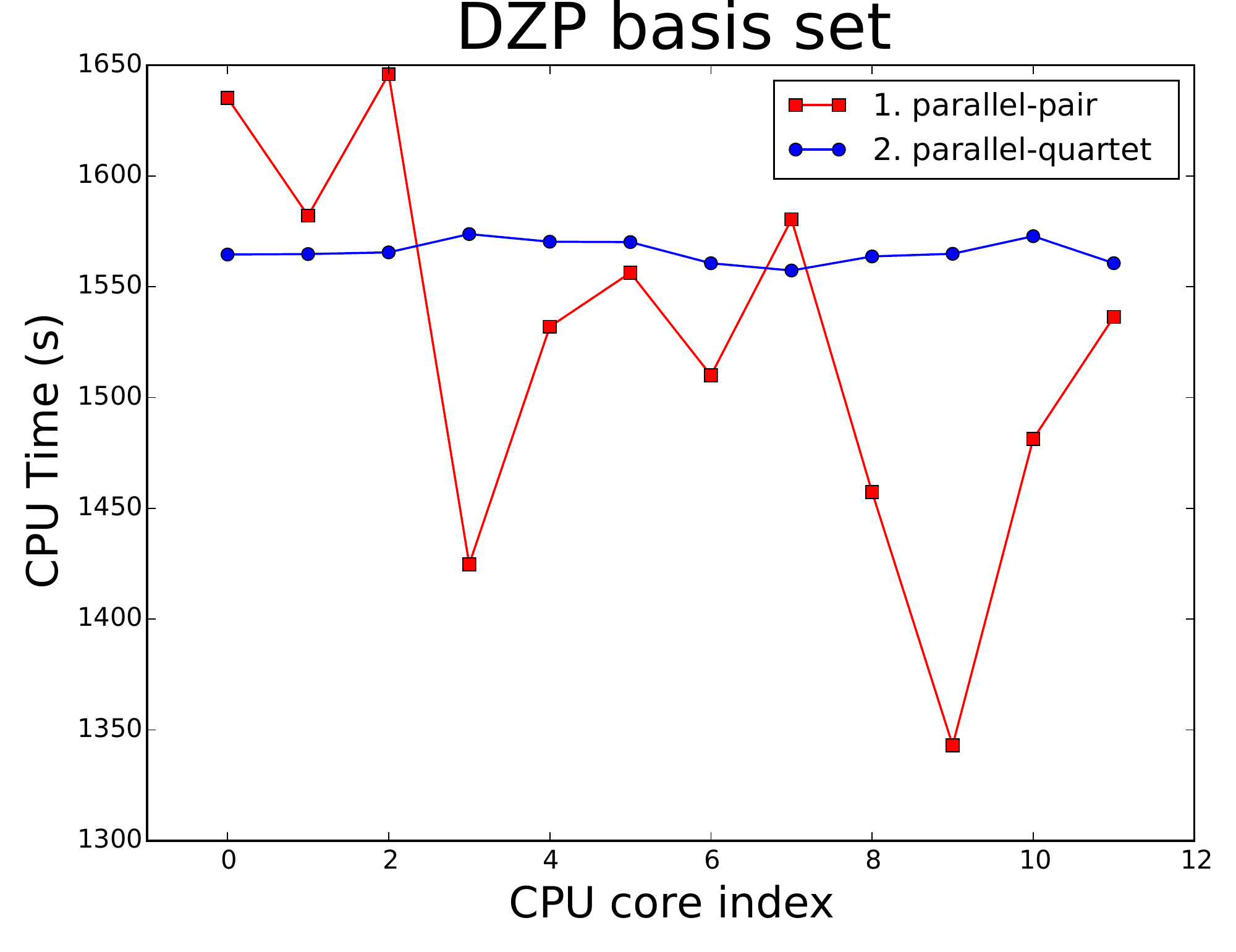}
\caption{The load balance for bulk silicon supercell with 64 atoms using DZP basis set at 12 CPU cores.}
\label{fig:si_64atoms_dzp_12core}
\end{figure}

\begin{figure}
\includegraphics[width=0.5\textwidth]{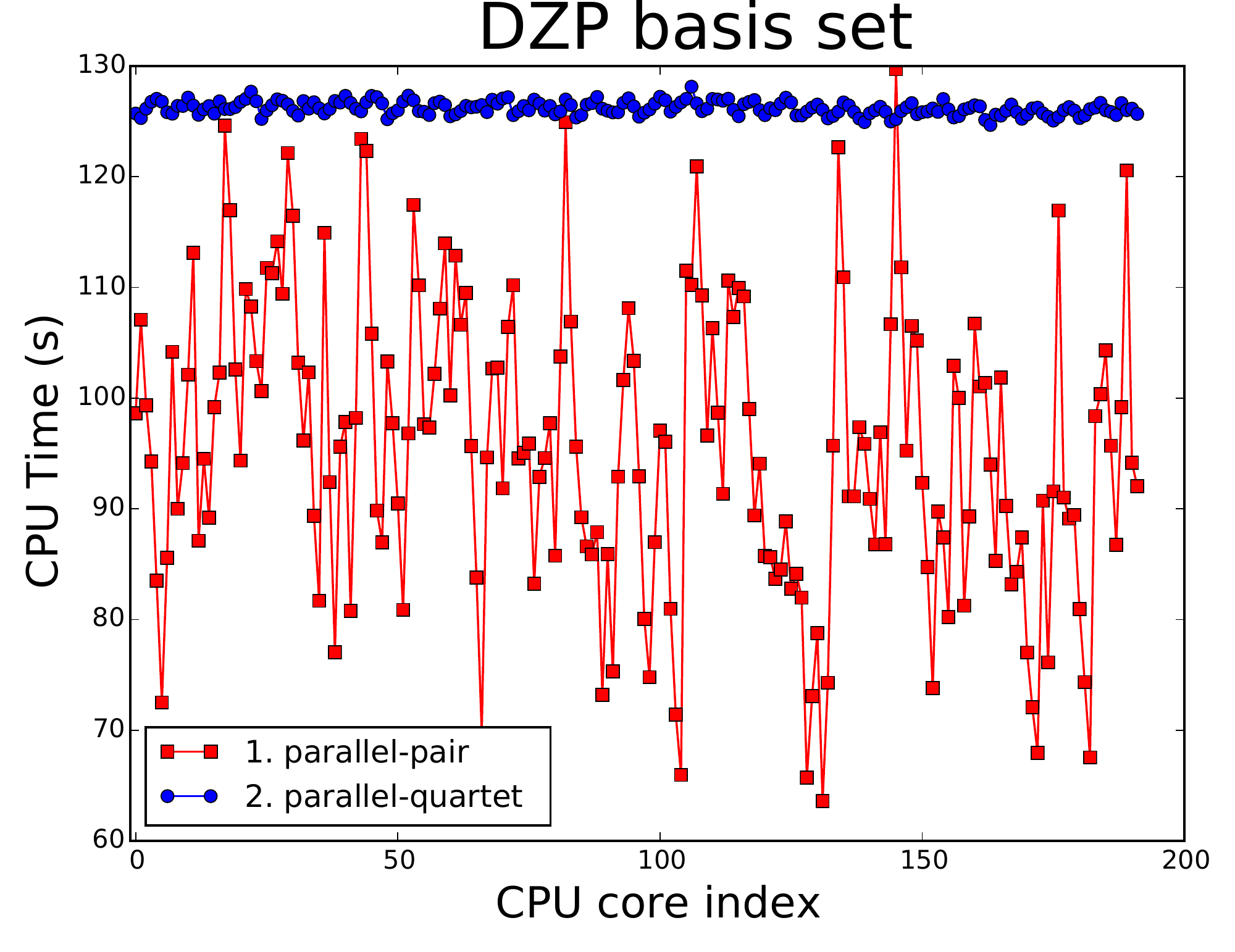}
\caption{The load balance for bulk silicon supercell with 64 atoms using DZP basis set at 192 CPU cores.}
\label{fig:si_64atoms_dzp_192core}
\end{figure}

For the parallel-pair and parallel-quartet algorithms, which are fully parallelized and involve no communication, load imbalance is one of factors that may affect the parallel efficiency. To examine the load balance, the  
timing at every cores are shown in Fig. \ref{fig:si_64atoms_sz_12core}-Fig. \ref{fig:si_64atoms_dzp_192core}. It is clearly
shown that the parallel-pair algorithm (red line) is 
load imbalance, for SZ basis set, the time difference between cores is around 10\% in 12 cores~(Fig. \ref{fig:si_64atoms_sz_12core}) case and around 80\% in 192 cores~(Fig. \ref{fig:si_64atoms_sz_192core}). For DZP basis set, d shells have been considered, which caused more serious load imbalance, the time difference between cores is even around 22\% in 12 cores~(Fig.\ref{fig:si_64atoms_dzp_12core}) and around 100\% in 192 cores~(Fig.\ref{fig:si_64atoms_dzp_192core}).
On the other hand, the load balance in parallel-quartet algorithm is quite well, the  time difference between cores is within 1\%. However, in this algorithm, as the ERIs screening part is made for the whole ERIs by all the CPU cores. which is a constant time even with the increasing CPU cores, there 
are replicate calculations for the ERIs screening which decrease the parallel efficiency. As shown in Fig.\ref{fig:si_64atoms_sz_192core}), for small basis set at large number of CPU cores, the average CPU time of the parallel-quartet 
is around twice as the parallel-pair algorithm.

\begin{table}
\caption{The CPU time~(in seconds) for the calculation of the ERIs of Si 64 system with SZ basis set using different parallel schemes. }
\label{tab:cpu_time_sz}
\begin{tabular}{c c c}
\hline
cores & parallel-pair & parallel-quartet \\
\hline
12    &    110.9        &    108.2             \\ 
24    &    55.6            &  57.0   \\
96    &     15.5          &   18.2   \\
192   &     8.6           &    11.6 \\
\hline
\end{tabular}
\end{table}

\begin{table}
\caption{The CPU time~(in seconds) for the calculation of the ERIs of Si 64 system with DZP basis set using different parallel schemes. }
\label{tab:cpu_time_dzp}
\begin{tabular}{c c c}
\hline
cores & parallel-pair & parallel-quartet \\
\hline
12    &    1645.1      &    1572.9             \\ 
24    &    904.0           &  806.1   \\
96    &     251.3          &   225.9   \\
192   &     129.6           &    128.0 \\
\hline
\end{tabular}
\end{table}

\begin{figure}
\includegraphics[width=0.5\textwidth]{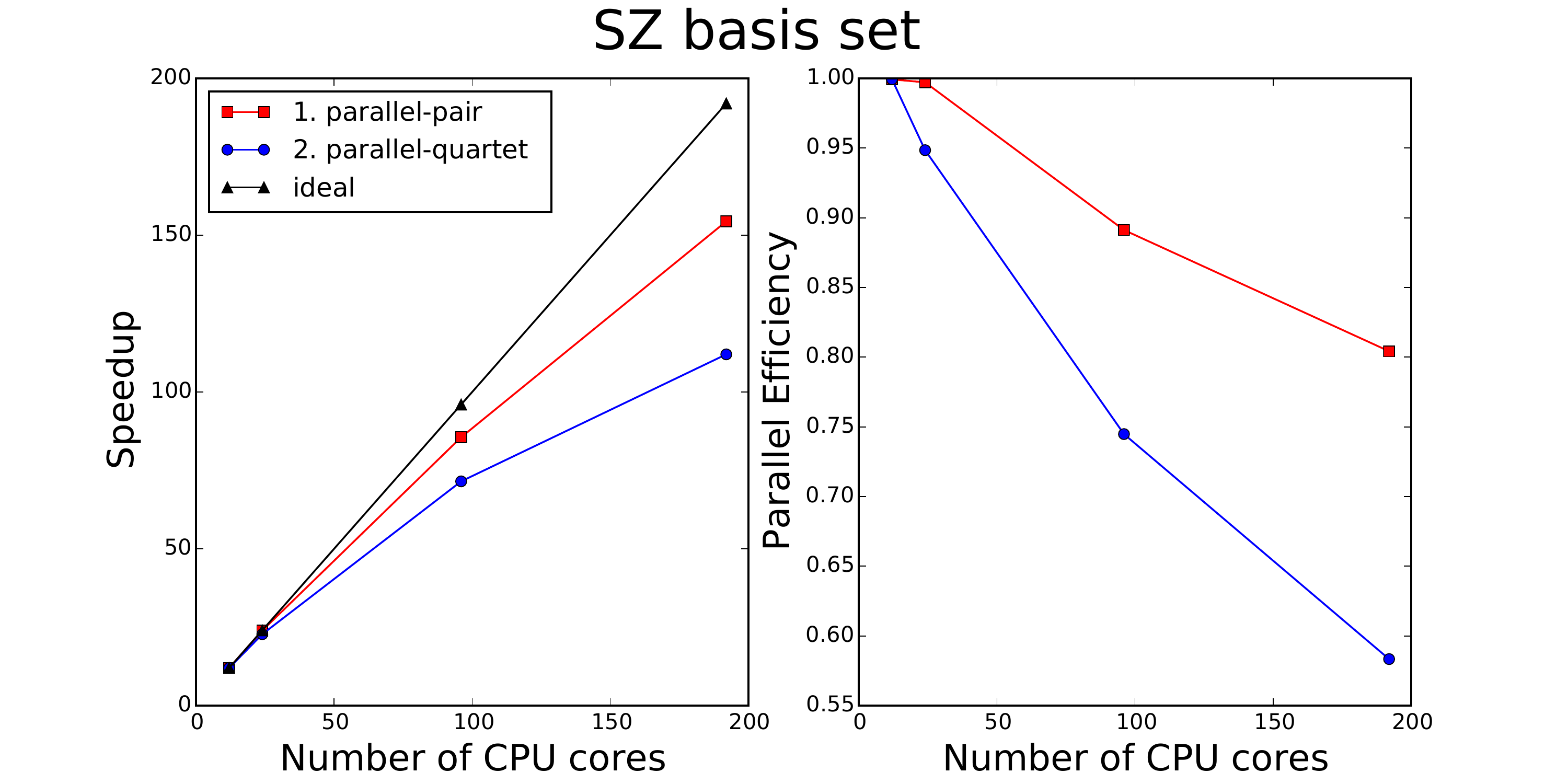}
\caption{ (Color online) Parallel Speedups and efficiency for ERIs calculation  formation using different parallel schemes.
Speedups were obtained on Tianhe-2 for bulk silicon supercell with 64 atoms using  SZ basis set.
The speedup is referenced to a run on 12 CPUs.}
\label{fig:speedup-sz}
\end{figure}

\begin{figure}
\includegraphics[width=0.5\textwidth]{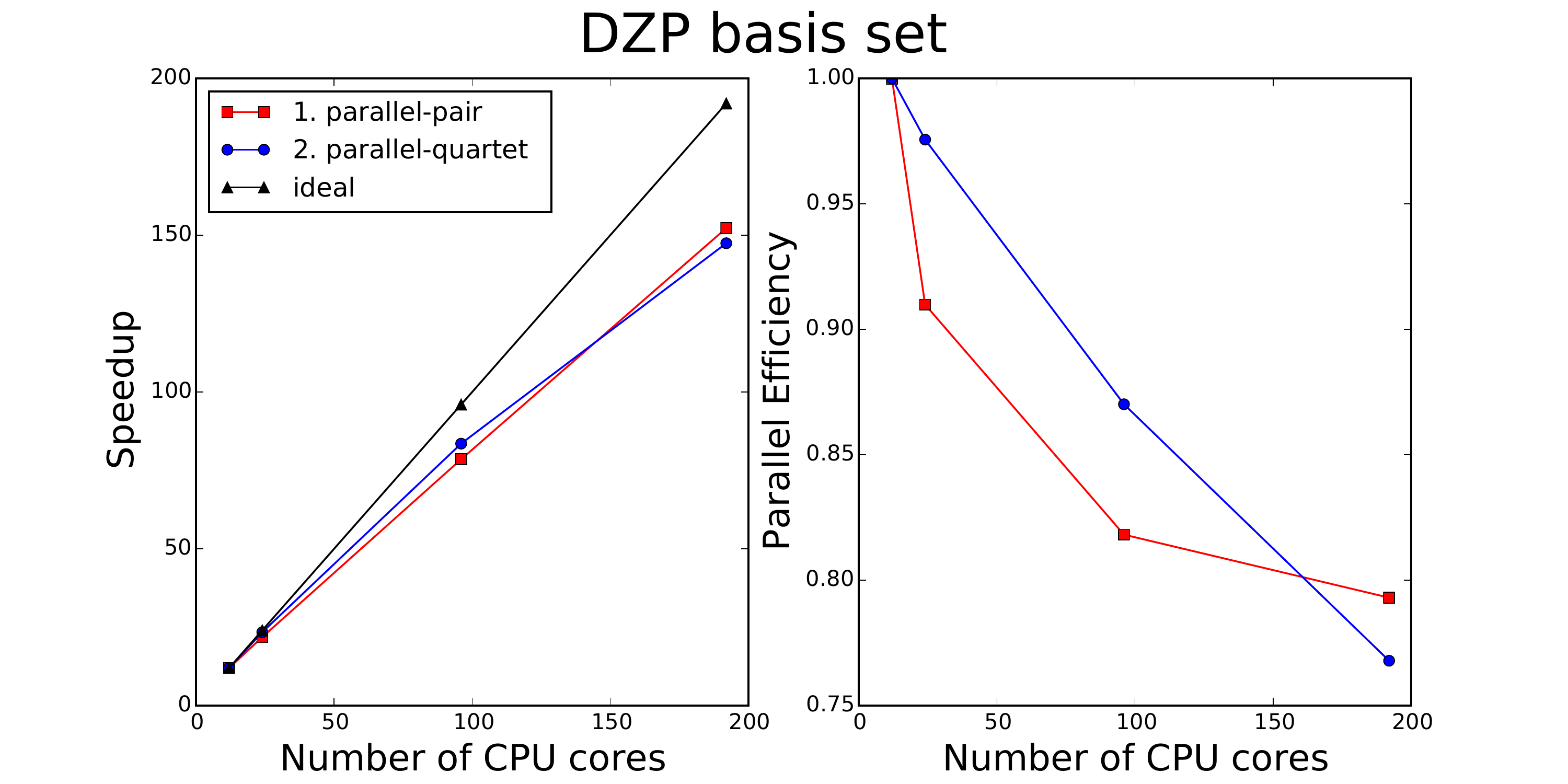}
\caption{ (Color online) Parallel Speedups and efficiency for ERIs calculation using different parallel schemes.
Speedups were obtained on Tianhe-2 for bulk silicon supercell with 64 atoms using  DZP basis set
The speedup is referenced to a run on 12 CPUs.}
\label{fig:speedup-dzp}
\end{figure}
Such global ERI screening in parallel-quartet algorithm also contributes significantly to lowering the parallel speedup and efficiency for SZ basis set, as shown in Table\ref{tab:cpu_time_sz} and Fig.\ref{fig:speedup-sz}. The 
parallel efficiency is only 58\% at 192 CPU core for parallel-quartet while holds around 80\% for parallel-pair algorithm. 

For DZP basis set, the load imbalance caused by d shells become another factor of lowering the parallel efficiency, so in this case, as shown in Table\ref{tab:cpu_time_sz} and Fig.\ref{fig:speedup-dzp}, the parallel speedup and efficiency are similar for both 
parallel-pair and parallel-quartet algorithms, that around 80\% at 192 CPU cores.

\section{CONCLUSIONS}
In summary, we have shown our two static parallel algorithms for the ERIs calculations in NAO2GTO method. We have also analyzed the performance of these two parallel algorithms for their load balance and parallel efficiency. On the basis of our results, 
the static distribution of ERI shell pairs, produces load imbalance that causes the efficiency to decrease, limiting the number of CPU cores that can be utilized. On the other hand, the static
distribution of ERI shell quartet can yield very high
load balance, however, because the need of the  global ERI screening calculation, the parallel efficiency has been dramatically reduced for small basis set.  We have also tried another static method that firstly create a need-to-calculate ERIs list by considering all the screening methods as well as the eight-fold permutational symmetry and secondly distribute the ERIs in the need-to-calculate list over a number of processes. However, we find the time to build the need-to-calculate ERIs list is even larger than the global ERI screening calculation.
On the next step, we need to distribute the ERI screening calculation while keep the load balance of the ERI calculation, and a dynamic distribution could enables load balance with little loss of efficiency.

\begin{acks}
This work was supported by the National Key Research $\&$ Development Program of China (2017YFB0202302), the Special Fund for Strategic Pilot Technology of Chinese Academy of Sciences (XDC01040000),  the National Natural Science Foundation of China (61502450), the National Natural Science Foundation of China (21803066), Research Start-Up Grants (KY2340000094) from University of Science and Technology of China, and Chinese Academy of Sciences Pioneer Hundred Talents Program. The authors thank the Tianhe-2 Supercomputer Center for computational resources.

\end{acks}

\bibliographystyle{SageH}
\bibliography{mybib}

\section{Author Biographies}
\textit{Xinming Qin} is a PhD student in chemical physics at USTC under the supervision of Prof.Jinlong Yang. He received the BS degree in chemistry in July 2009 at the same university. His main research interests are developing and applying new algorithms for large-scale electronic structure calculations. He is the main developer of the HONPAS.

\textit{Honghui Shang} is an associate professor at the Computer Science at Institute of Computing Technology, Chinese Academy of Sciences, China. She received her BS degree in physics from University of Science and Technology of China in 2006, and the PhD degree in Physical Chemistry from University of Science and Technology of China, in 2011. Between 2012 and 2018, she worked as a postdoctoral research assistant at the Fritz Haber Institute of the Max Planck Society, Berlin, Germany, which was responsible for the Hybrid Inorganic/Organic Systems for Opto-Electronics (HIOS) project and for the Novel Materials Discovery (NOMAD) project. Her main research interests are developing the physical algorithm or numerical methods for the first-principle calculations as well as accelerating these applications in the high performance computers. Currently, she is the main developer of the HONPAS (leader of of the hybrid-density-functional part) and FHI-aims~(leader of the density-functional perturbation theory part).

\textit{Lei Xu} is currently working toward the BS degree in the department of physics at the SiChuan University. His current research interests are the parallel programming in the high performance computing domain.

\textit{Wei Hu} is currently a professor at division of theoretical and computational sciences at Hefei National Laboratory for Physical Sciences at Microscale (HFNL) at USTC. He received the BS degree in chemistry from USTC in 2007, and the PhD degree in Physical Chemistry from the same university in 2013. From 2014 to 2018, he worked as a postdoctoral fellow in the Scalable Solvers Group of the Computational Research Division at Lawrence Berkeley National Laboratory (LBNL), Berkeley, USA. During his postdoctoral research, he developed a new massively parallel methodology, DGDFT (Discontinuous Galerkin Method for Density Functional Theory), for large-scale DFT calculations. His main research interests focus on development, parallelization, and application of advanced and highly efficient DFT methods and codes for accurate first-principles modeling and simulations of nanomaterials.

\textit{Jinlong Yang} is a full professor of chemistry and executive dean of the School of Chemistry and Material Sciences at USTC. He obtained his PhD degree in 1991 from USTC. He was awarded Outstanding Youth Foundation of China in 2000, selected as Changjiang Scholars Program Chair Professor in 2001 and as a fellow of American Physical Society (APS) in 2011. He is the second awardee of the 2005’s National Award for Natural Science (the second prize) and the awardee (principal contributor) of the 2014’s Outstanding Science and Technology Achievement Prize of the Chinese Academy of Sciences (CAS). His research mainly focuses on the development of first-principles methods and their application to clusters, nanostructures, solid materials, surfaces, and interfaces. He is the initiator and leader of the HONPAS.

\textit{Shigang Li} received his Bachelor in computer science and technology and PhD in computer architecture from the University of Science and Technology Beijing, China, in 2009 and 2014, respectively. He was funded by CSC for a 2-year PhD study in University of Illinois, Urbana-Champaign. He was an assistant professor (from June 2014 to Aug. 2018) in State Key Lab of Computer Architecture, Institute of Computing Technology, Chinese Academy of Sciences at the time of his achievement in this work. From Aug. 2018 to now, he is a postdoc researcher in Department of Computer Science, ETH Zurich. His research interests focus on the performance optimization for parallel and distributed computing systems, including parallel algorithms, parallel programming model, performance model, and intelligent methods for performance optimization.

\textit{Yunquan Zhang} received his BS degree in computer science and engineering from the Beijing Institute of Technology in 1995. He received a PhD degree in Computer Software and Theory from the Chinese Academy of Sciences in 2000. He is a full professor and PhD Advisor of State Key Lab of Computer Architecture, ICT, CAS. He is also appointed as the Director of National Supercomputing Center at Jinan and the General Secretary of China's High Performance Computing Expert Committee. He organizes and distributes China's TOP100 List of High Performance Computers, which traces and reports the development of the HPC system technology and usage in China. His research interests include the areas of high performance parallel computing, focusing on parallel programming models, high-performance numerical algorithms, and performance modeling and evaluation for parallel programs.

\end{document}